\input{aipcheck}
\newcommand{\rem}[1]{ }
\newcommand{\beq}{\begin{equation}}
\newcommand{\eeq}{\end{equation}}
\newcommand{\bea}{\begin{eqnarray}}
\newcommand{\eea}{\end{eqnarray}}
\renewcommand{\cite}[1]{\citep{#1}}
\def\apj{{ApJ}}
\def\apjl{{ApJL}}

\def\prl{{PRL}}

\documentclass[
    ,final            
  ]
  {aipproc}

\layoutstyle{6x9}

\begin{document}

\title{Physics of relativistic shocks}


\author{Mikhail V. Medvedev}{
  address={Department of Physics and Astronomy, 
University of Kansas, KS 66045}
}

\begin{abstract}
Relativistic shocks are usually thought to occur in violent astrophysical explosions. These collisionless shocks are mediated by a plasma kinetic streaming instability, often loosely referred to as the Weibel instability, which generates strong magnetic fields ``from scratch" very efficiently. In this review paper we discuss the shock micro-physics and present a recent model of ``pre-conditioning" of an initially unmagnetized upstream region via the cosmic-ray-driven Weibel-type instability. 
\end{abstract}

\maketitle

\section{Introduction}

It has been shown in recent years that collisionless relativistic 
shocks are mediated by the Weibel instability --- a current filamentation 
instability that produces strong, sub-equipartition magnetic fields at the 
shock front \citep{ML99}. This is an attractive model 
for gamma-ray bursts (GRBs),
because it puts a synchrotron shock model on a firm physical ground.
Here we describe the physics of the instability, its linear regime and saturation. We discuss the properties of the nonlinear Weibel turbulence --- the state of the ongoing self-similar process of current mergers. Finally, we present a toy model of the foreshock region --- an extended region in the upstream of the shock which may likely be populated with strong and relatively large-scale (meso-scale) magnetic fields produced by shock-accelerated particles. The foreshock can be the main player in determining the shock radiative processes and the efficiency of cosmic ray acceleration.

\section {Linear regime of field growth and its saturation}

The instability under consideration was first predicted by \citet{Weibel59}
for a non-relativistic plasma with an anisotropic distribution function. 
The simple physical interpretation considers the PDF anisotropy more 
generally as a two-stream configuration 
of cold plasma. Below we give a brief, qualitative description of this 
streaming magnetic instability.

Let us consider, for simplicity, the dynamics of one species only 
(e.g., protons), whereas the other (electrons) is assumed to provide global charge 
neutrality.\footnote{In reality, the role of protons is more complicated, e.g., 
they play a crucial role in the electron heating, which we do not consider here.}
The streaming particles are assumed to move along the $z$-axis with the velocities 
${\bf v}=+\hat z v_z$  and ${\bf v}=-\hat z v_z$, thus forming  
equal particle fluxes in opposite directions (so that 
the net current is zero). Such a particle distribution occurs naturally
near the front of a shock (moving along $z$-direction), 
where the ``incoming'' (in the shock frame) 
ambient gas particles meet the ``outgoing'' particles reflected from the
shock potential (loosely speaking, the low energy cosmic rays). 
Thus, the particle velocities $v_z$ are of order the shock velocity,
$v_z\sim v_{\rm sh}$.
The counter-streaming particles may also have some thermal 
spread. Since for high-Mach number shocks, $v_{\rm th}\ll v_{\rm sh}$ in the 
upstream region, we may neglect the parallel velocity spread in our 
consideration. The thermal spread in the transverse direction cannot be
neglected, however. We parameterize the PDF anisotropy
as follows:
\beq
A=\frac{\epsilon_\|-\epsilon_\bot}{\epsilon_{\rm tot}}
\simeq\frac{M^2-1}{M^2+1},
\eeq
where $\epsilon_\|\propto v_z^2\simeq v_{\rm sh}^2$ 
is the energy of particle along $z$-direction,
$\epsilon_\bot\propto (v_x^2+v_y^2)\propto v_{\rm thermal}^2\simeq c_s^2$ 
is the thermal energy in the transverse direction,
$\epsilon_{\rm tot}=\epsilon_\|+\epsilon_\bot$ is the total energy,
$c_s$ is the sound speed upstream and the Mach number of the shock is
$M=v_{\rm sh}/c_s$. Clearly, for strong shocks $M\gg1$, the anisotropy 
parameter is close to unity, $A\sim1$.
Next, according to the linear stability analysis
technique, we add an infinitesimal magnetic field fluctuation, 
${\bf B}=\hat x B_x \cos(ky)$. The Lorentz force, $ e({\bf v \times B})/c$,
acts on the charged particles and deflects their trajectories,
as is shown in Figure 1a.
\begin{figure}
$\begin{array}{l}
  \includegraphics[height=.3\textheight]{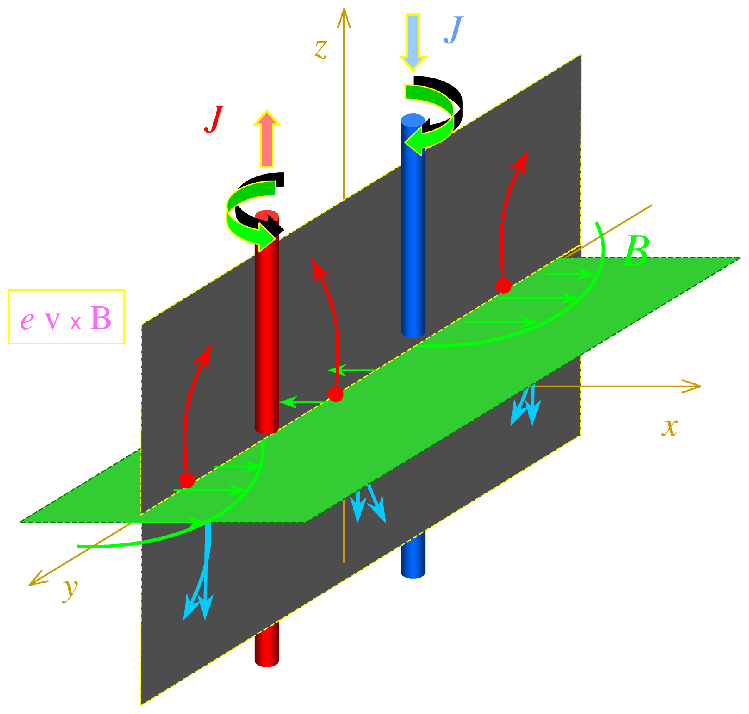}(a)\\
  \includegraphics[height=.3\textheight]{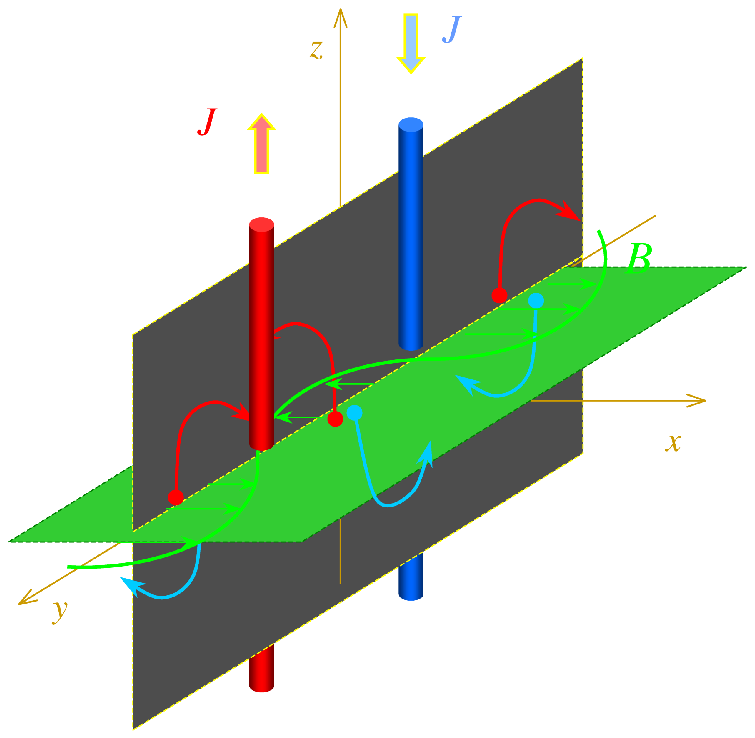}(b)\\
  \includegraphics[height=.3\textheight]{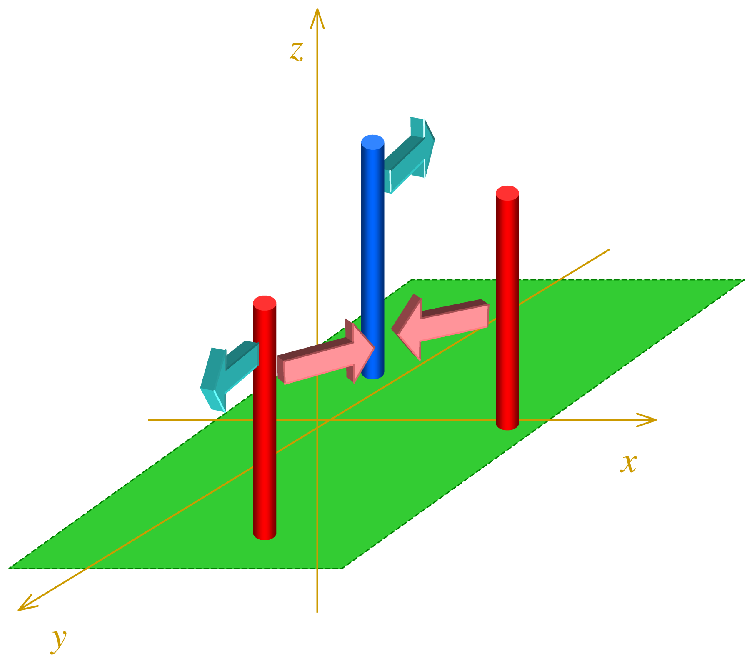}(c)
  \end{array}$
  \caption{Illustration of various stages of the Weibel instability.
Color coding of particles: {\it blue} --- the incoming particles from the IGM,
{\it red} --- the particles scattered from the shock. 
(a) Linear regime: current filamentation; (b) saturation; 
(c) nonlinear regime: filament coalescence.}
\label{f:1}
\end{figure}
As a result, the protons moving upward and those moving downward
will concentrate in spatially separated current filaments. 
The magnetic field of these filaments appears to increase the 
initial magnetic field fluctuation. The growth rate and the wavenumber of
the fastest growing mode (which, in fact, sets the spatial correlation
scale of the produced field) are
\beq
\gamma_B=A\,\omega_{p,s} (v_z/c), \qquad k_B=A\,\omega_{p,s}/c,
\eeq
where 
$
\omega_{p,s}=\left({4\pi e^2 n_s}/{m_s}\right)^{1/2}
$
is the plasma frequency defined for species $s$ (electrons, protons, etc.),
$n_p$ and $m_p$ are the number density and the mass of the protons,
respectively. The above scalings are for non-relativistic plasmas. In the relativistic case, one shall simply substitute $m_s\to\gamma_s m_s$, there $\gamma_s$ is the Lorentz factor of the streaming species. 
Note that the instability is driven by the PDF anisotropy and should 
quench for the isotropic case, $A=0$. 

The Lorentz force deflection of particle orbits increases as the 
magnetic field perturbation grows in amplitude. The amplified magnetic 
field is random in the plane perpendicular to the particle motion, 
since it is generated from a random seed field. Thus, the Lorentz 
deflections result in a pitch angle scattering, which makes the bulk of the 
PDF isotropic. If one starts from a strong anisotropy, so that the thermal 
spread is much smaller than the particle bulk velocity, most of the particles 
will eventually isotropize and the thermal energy associated with their
random motions will be equal to their initial directed kinetic energy. 
This final state will bring the instability to saturation.

The saturation level of the magnetic field may readily be estimated as 
follows.  First of all, note that the instability is due to the free 
streaming of particles. As the magnitude of the magnetic field grows, 
the transverse deflection of particles gets stronger, and their free 
streaming across the field lines is suppressed, see Figure 1b. Once the bounce rate of the streaming particles in the self-generated current filaments becomes greater than the instability growth rate, the instability ceases. In the relativistic case we are concerned with here, this condition is identical to that  
the typical curvature scale for the deflections is the Larmor radius, 
\beq
\rho_L=v_{\bot B}/\omega_{c,s},
\eeq
where $v_{\bot B}$ is the 
particle velocity transverse to the direction of the local magnetic field and 
\beq
\omega_{c,s}=\frac{eB}{\gamma_s m_s c}
\eeq
is the cyclotron (Larmor) frequency of species $s$ and we assumed $v_{\bot B}\sim c$. 
On scales larger than $\rho_L$, particles can only move
along field lines. Hence, when the growing magnetic fields become such that
$k_{B}\rho_L\sim1$, the particles are magnetically trapped and can no
longer amplify the field. Assuming an isotropic particle distribution at
saturation ($v_{\bot B}\sim v_{\rm sh}$), this condition can be re-written as 
\beq 
\epsilon_B=\frac{B^2/8\pi}{m_s c^2 n_s \gamma_{shock}^2 } \simeq A^2 .
\eeq 
For strong shocks ($M\gg1,\ A\sim 1$), this corresponds to the 
magnetic energy density close to equipartition with the thermal energy
of particles downstream the shock. 

Numerical PIC simulation indicate a somewhat smaller value of $\epsilon_B$, typically not exceeding 0.1 or so even for strong shocks, that is the magnetic field energy density at a relativistic shock is of the order of 10\% of the shock kinetic energy \citep{Silva+03,Fred+04,Spit08}. For more recent relevant work, see \citep{w1,w2}.

\section {Nonlinear Weibel turbulence and filament coalescence}

Numerous 3D PIC numerical simulations demonstrate that the
generated magnetic fields are associated with a quasi-two-dimensional 
distribution of current filaments. Hence we suggest the
following toy model. We consider straight one-dimensional current filaments oriented in the
vertical, $z$-direction. Initially, all filaments are identical:
the initial diameter of them is $D_0$, their initial mass per unit length is
$\mu_0\simeq mn(\pi D_0^2/4)$, where $m$ is the mass of plasma particles 
(e.g., electrons) and $n$ is their number density. Each filament carries
current $I_0$ in either positive or negative $\hat z$-direction.
The net current in the system is set to zero, i.e., there are equal numbers
of positive and negative current filaments. We also assume that
the initial separation (the distance between the centers) 
of the filaments $d_0$ is comparable to their size, $d_0\simeq 2D_0$.
Finally, no external homogeneous magnetic field is present in the system.
Here we assume that filaments have a simple structure: they have no return current on the outside. Generally, this is not true, as the filaments represent some sort of the Harris equilibrium. The return current can be incorporated in the analysis and is expected to decrease the overall coalescence rate, because it shall partially screen the main current and, thus, reduce the interaction strength between the filaments. 

Initially, the filaments are at rest and their positions in space are random.
This configuration is unstable because opposite currents repel each other,
whereas like currents are attracted to each other and tend to 
coalesce and form larger current filaments.
The characteristic scale of the magnetic field will accordingly
increase with time. We study this process quantitatively using
the toy model of two interacting filaments. 

The magnetic field produced by a straight filament is $B_0(r)=2I_0/(cr)$, 
where $r$ is the cylindrical radius. The force per unit length acting on 
the second filament is $dF/dl=-B_0I_0/c$. Since $dF/dl=\mu \ddot x$, where
$x$ is the position in the center of mass frame and ``overdot'' denotes 
time derivative, we write the equation of motion as follows:
\beq
\ddot x=-\frac{2I_0^2}{c^2\mu_0}\,\frac{1}{x},
\label{eom}
\eeq
where we used that $r=2x$ and the reduced mass $\mu_r=\mu_0/2$.  
We define the coalescence time as the time required for the 
filaments, which are initially at rest, to cross the distance between 
them and ``touch'' each other, which happens when the distance
between their centers becomes equal to $D_0$, i.e., when $x=D_0/2$.
The coalescence time, as it is defined above, is independent of the details
of the merging process itself, which involves rather complicated 
dynamics associated with the redistribution of currents.
Quite obviously, the interaction between the filaments is the weakest
at large distances $x\sim x_0\sim d_0/2$. Hence, the coalescence rate 
is limited by the filament motions at the largest scales. 
The coalescence time can 
be readily estimated from Eq. (\ref{eom}), assuming that 
$x\sim x_0\sim d_0/2$ and $\ddot x\sim (d_0/2)\tau_0^{-2}$, as follows:
\beq
\tau_{0,NR}\sim\left({D_0^2 c^2 \mu_0}/({2 I_0^2})\right)^{1/2}.
\label{tau0NR}
\eeq 
The above estimate is valid as long as the motion is non-relativistic. 
The maximum velocity of a filament is at the time of coalescence, $x=D_0/2$:
\beq
v_{0}\sim D_0/2\tau_0\sim I_0/(c\sqrt{2\mu_0}). 
\label{vm0}
\eeq
It must always be much smaller than the speed of light.

If the motion of a filament during the merger becomes sub-relativistic, 
i.e., $v_{merger}$ becomes a significant fraction of $c$
the separation cannot decrease faster than as
$t(x)\simeq x/c$.
Therefore, the coalescence time will be
\beq
\tau_{0,R}\simeq  D_0/v_{merger}.
\label{tau0R}
\eeq

The filament coalescence is a hierarchical process.
Indeed, suppose that initially the system contains $N_0$ current filaments,
with an average separation $d_0\sim 2D_0$. Each of the filaments 
carries current $I_0$, its diameter is $D_0$ and its mass per unit
length is $\mu_0$. For simplicity, we assume that filaments coalesce
pairwise. 

Having the original ``zeroth  generation'' of filaments merged 
(the process takes about $\tau_{0,NR}$ or $\tau_{0,R}$ to complete),
the system will now contain $N_0/2$ of ``first generation'' filaments.
Each of these filaments carries current $I_1=2I_0$, has mass per
unit length $\mu_1=2\mu_0$, and the separation between them 
is $d_1=\sqrt{2}d_0$ (because the two dimensional number density of filaments 
decreased by 2). Since $\mu\propto D^2$, the filament size also increases
as $D_1=\sqrt{2} D_0$. Remarkably, this new configuration is  
identical to the initial one, but with the re-scaled parameters. 
Hence, the coalescence process is self-similar. 
The produced first generation  filaments will be interacting with 
each other and merge again to yield the second generation.
The coalescence process will then continue in a self-similar way.
Note that the coalescence times at each stage are not necessarily the same.
Taking into account that at the $k$-th merger level, i.e., 
after $k$ pairwise mergers:
$I_k=2^k I_0,\  \mu_k=2^{k} \mu_0,\  D_k=2^{k/2} D_0,\ d_k\sim D_k/2$, 
we obtain
\beq
\tau_{k,NR}= \tau_{0,NR},  \qquad
\tau_{k,R}= 2^{k/2}\tau_{0,R}. \label{tau-k}
\eeq
Since the coalescence time is independent of $k$ while the filaments 
are non-relativistic, whereas the distance between them increases, 
the typical velocities of the merging filaments grow with time and,
will approach some terminal velocity $v_{merger}\sim c$, as $v_{k}=2^{k/2}v_{0}$.

Finally, it is instructive to present the evolution of the parameters
as a function of physical time, $t$, rather than the merger level, $k$.
Apparently, it takes $t=\sum_{k'=0}^k\tau_{k'} $
to complete $k$ mergers, where $\tau_k$ is given by Eq. (\ref{tau-k}). 
Thus, for the non-relativistic and relativistic cases respectively, 
we have: $k={t}/{\tau_{0,NR}}$ and 
$k\simeq2\log_2\left[({t}/{\tau_{0,R}})\right]$. 
Thus, the magnetic field correlation length increases as a function of
time as
\beq
\lambda_B(t)=D_0 2^{{t}/({2\tau_{0,NR}})}, \qquad
\lambda_B(t)\simeq  v_{merger}t, \label{lambda}
\eeq
in the non-relativistic and sub-relativistic regimes, respectively. Note that
the last expression is an approximation at large times $t\gg\tau_{0,R}$,
i.e., at large $k\gg1$. 

\begin{figure}
  \includegraphics[height=.2\textheight]{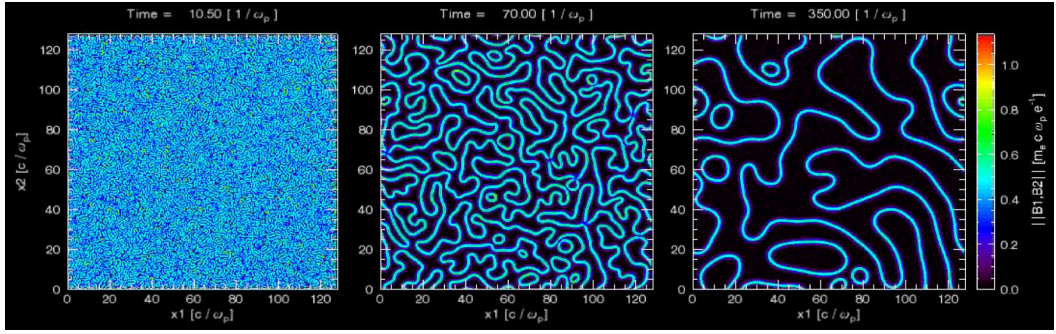}
\caption{The 2D structure of the magnetic fields in $e^-e^+$
plasmas at various times (from \citep{M+04}).
The change of the field correlation length with time is clearly seen.
Similar The growth of this length is substantially slower and the magnetic field 
filling factor is respectively larger in the electron-proton plasma.  }
\label{f:3}
\end{figure}

Simulations of the Weibel filament dynamics in 2D pair-dominated and electron-ion plasmas confirm the above arguments. A few snapshots are shown in Figure \ref{f:3}.
It has been observed that the filament coalescence does occur and the merger velocity in relativistic plasmas does approach some terminal value, which is somewhat smaller then the speed of light. By determining the spectral peak of the magnetic fluctuations at each time step, one determined that after an initial fast growth, the typical correlation length of the field sets on the scalings
\beq
\lambda_B(t)\propto t^{0.8}, \qquad
\lambda_B(t)\propto t, 
\eeq
for non-relativistic and relativistic plasmas, respectively. Thus, the field scale grows in a self-similar manner, as one shall expect for a scale-free turbulence.
We also note that in some respect, the field scale growth is analogous to the 
inverse cascade in two-dimensional magnetohydrodynamic (MHD) turbulence. 
The crucial difference is, however, the 
entirely {\em kinetic} nature of the process; at such small scales
$\sim c/\omega_p$ the MHD approximation is completely inapplicable.

\section{Relativistic shock and its foreshock region}

The simple understanding of a relativistic shock goes as follows. The particles reflected from the shock potential and propagate forward. They are Weibel-unstable and result in plasma current filamentation. The filaments interact with each other; they merge to form larger ones with stronger magnetic field since the current is approximately conserved during mergers. As filaments get stronger, they scatter particle more, thus causing particle heating and acceleration. At some point, the filaments are too strong, so they break up and result in strong longitudinal electric fields that slow down the ions (or the plasma in-flow, in general). Here the density builds up quickly, so this is the shock jump. Such a picture is, however, rather simplistic. The shock continuously accelerates and ejects cosmic rays (CRs) into the upstream region. Because of the non-linear shock acceleration, these CRs carry a substantial fraction of the shock energy, yet they are propagating nearly at the light speed -- much faster than the shock. Hence, in time, the CRs deposit their energy into the magnetic field in an extremely large region in front. These fields will in turn affect the shock propagation and particle acceleration. The model of the relativistic foreshock is discussed now.

Overall, our model of the foreshock is as follows. A shock is a source of CRs which move away from it, thus forming a stream of particles through the ambient medium, say, the interstellar medium (ISM). If the ISM magnetic fields are negligible, i.e., their energy density is small compared to that of CRs, the streaming instability (either the pure magnetostatic Weibel or the mixed-mode electromagnetic oblique Weibel-type instability, depending on conditions) is excited and stronger magnetic fields are quickly generated. The Weibel instability dispersion equation \citep{Weibel59} has been also derived as the cosmic-ray streaming instability dispersion equation, see e.g.,  \citep{Lee82}. The generated fields further isotropize (thermalize) the CR stream. Since less energetic particles, having a greater number density and carrying more energy overall, are thermalized closer to the shock, the generated B-field will be stronger closer to the shock and fall off away from it, whereas its correlation length will increase with the increasing distance from the shock.  More energetic particles keep streaming because of their larger Larmor radii and produce the magnetic field further away from the shock. This process stops at distances where either the CR flux starts to decrease (because of the finite distance the CR particles can get away from a relativistic shock or because of the shock curvature causing CR density to decrease as $\propto r^{-2}$ if the shock is sub- or non-relativistic) or where the generated magnetic fields become comparable to the ISM field and the instability ceases. Thus, a large upstream region --- the foreshock --- is populated with magnetic fields. We now derive its self-similar structure. We work in the shock co-moving frame unless stated otherwise. 

Let's consider a relativistic shock moving along $x$-direction with the bulk Lorentz factor $\Gamma_{\rm sh}$; the shock is plane-parallel and lies in the $yz$-plane, and $x=0$ denotes the shock position. The shock continuously accelerates cosmic rays, which then propagate away from it into the upstream region. We conventionally assume that the CR distribution over the particle Lorentz factor is described by a power-law: 
\beq
n_{\rm CR}=n_0(\gamma/\gamma_0)^{-p+1}
\label{nCR}
\eeq
for $\gamma>\gamma_0$ and zero otherwise. The index $p$ is approximately equal to 2.2 for ultrarelativistic shocks and $n_0$ is the normalization.\footnote{Conventionally the distribution is given as $dn/d\gamma\propto\gamma^{-p}$ with $p$ being $\sim2.2-2.3$ for relativistic shocks; hence the density of particles of energy $\sim\gamma$ is $n(\gamma)\propto\gamma^{-p}\delta\gamma\propto\gamma^{-p+1}$.}
 We assume that the above energy distribution is the same everywhere in the upstream, that is, we neglect the nonlinear feedback of magnetic fields onto the particle distribution. 
 Moreover, we assume that the formation of the CR power-law distribution is co-temporaneous with the shock formation itself, so we neglect the finite time acceleration of CRs, which may be quite important for higher-energy CRs. Accurate inclusion of this effect would require solution of the convection-diffusion equation with diffusion being calculated self-consistently from the self-generated fields, which are not steady at the beginning of the shock formation; all these issues are beyond the scope of the present paper.
 The CR momentum distribution exhibits strong anisotropy: the parallel ($x$) components of CR momenta are much greater than their thermal spread in the perpendicular ($yz$) plane. Indeed, for a particle to move away from the shock, it should have the $x$-component of the velocity exceeding the shock velocity. Since both the shock and the particle move nearly at the speed of light, this puts a constraint on their relative angle of propagation to be less then $1/\Gamma_{\rm sh}$ in the lab (observer) frame. Hence, the transverse spread of the CR particle's momenta is $p_\perp\le p_\|/ \Gamma_{\rm sh} \ll p_\|$. This is also seen in numerical simulations \citep{Spit08}.

The CR particles propagate through the self-generated foreshock fields and scatter off them. Lower energy particles are deflected in the fields more strongly and, therefore, izotropize faster than the higher energy ones, as having larger Larmor radii.  At a position $x>0$ the CR distribution can roughly be divided into isotropic (themalized) component with $\gamma<\gamma_r(x)$ and streaming component with $\gamma>\gamma_r(x)$, where $\gamma_r(x)$ is the minimum Lorentz factor of the streaming particles at a location $x$; it is also the maximum Lorentz factor of the randomized component at this location. The streaming component is Weibel-unstable with a very short $e$-folding time $\tau\sim\omega_{p,{\rm rel}}^{-1}$, where $\omega_{p,{\rm rel}}=\left(4\pi e^2 \tilde n(\gamma)/m_p \gamma\right)^{1/2}$ is the relativistic plasma frequency, $\tilde n(\gamma)$ is the density of streaming particles of the Lorentz factor $\gamma$ (tilde denotes streaming particles). Note that the Weibel instability growth rate depends on $n$ of the lower density component -- cosmic rays, in our case -- measured in the center of mass frame of the streaming plasmas. For the lower-energy part of the CR distribution, the center of mass frame is approximately the shock co-moving frame, hence we evaluate the instability on the shock frame. This approximation is less accurate for the high-energy CR tail; however, the growth rate and the scale length are weak functions of the the shock Lorentz factor ($\propto\Gamma_{\rm sh}^{\mp 1/2}$), so the result will be accurate within an order of magnitude for all reasonable values of $\Gamma_{\rm sh}$ for GRB afterglows.  Here we use the proton plasma frequency because the CR electron Lorentz factors are about $m_p/m_e$ times larger, so they behave almost like protons \citep{Spit08}. The instability is very fast: it rapidly saturates (the fields cease to grow) in a few tens of $e$-folding times $\tau$, that is in few tens of inertial  lengths (also referred to as the ion skin length)  $c/\omega_{p,{\rm rel}}$ in front of the shock.  Thereafter the particles keep streaming in current filaments and the field around them amounts to $\xi_B\sim0.01-0.001$ or so of the kinetic energy of this group of particles: 
\beq
B^2(\gamma)/8\pi\sim\xi_Bm_pc^2\gamma \tilde n(\gamma),
\label{B-gamma}
\eeq 
where $\xi_B$ is the efficiency factor obtained from PIC simulations; $\xi_B$ has the same meaning as the conventional $\epsilon_B$ parameter reserved here for the ratio of the total magnetic energy to the total kinetic energy of the shock and which, as is seen in PIC simulations, is larger than $\xi_B$ near the shock because of the nonlinear evolution and filament mergers. Whereas there is a concern that the Weibel instability can be suppressed by large upstream magnetic fields, numerical simulations generally indicate that the upstream field has little influence on the Weibel instability if the field energy density does not exceed few percent of the total shock kinetic energy density. For our conditions with $\xi_B\sim0.01-0.001$ and reasonable CR acceleration efficiency, $\xi_{\rm CR}\sim0.5$ (see, e.g., \citep{V+06}) the instability is hardly suppressed.
The correlation length of the field is of the order of the ion inertial (skin) length 
\beq
\lambda(\gamma)\sim c/\omega_{p,{\rm rel}}=\left(m_pc^2 \gamma/4\pi e^2 \tilde n(\gamma)\right)^{1/2}.
\label{lambda-gamma}
\eeq

\begin{figure}
\includegraphics[width=5.5in]{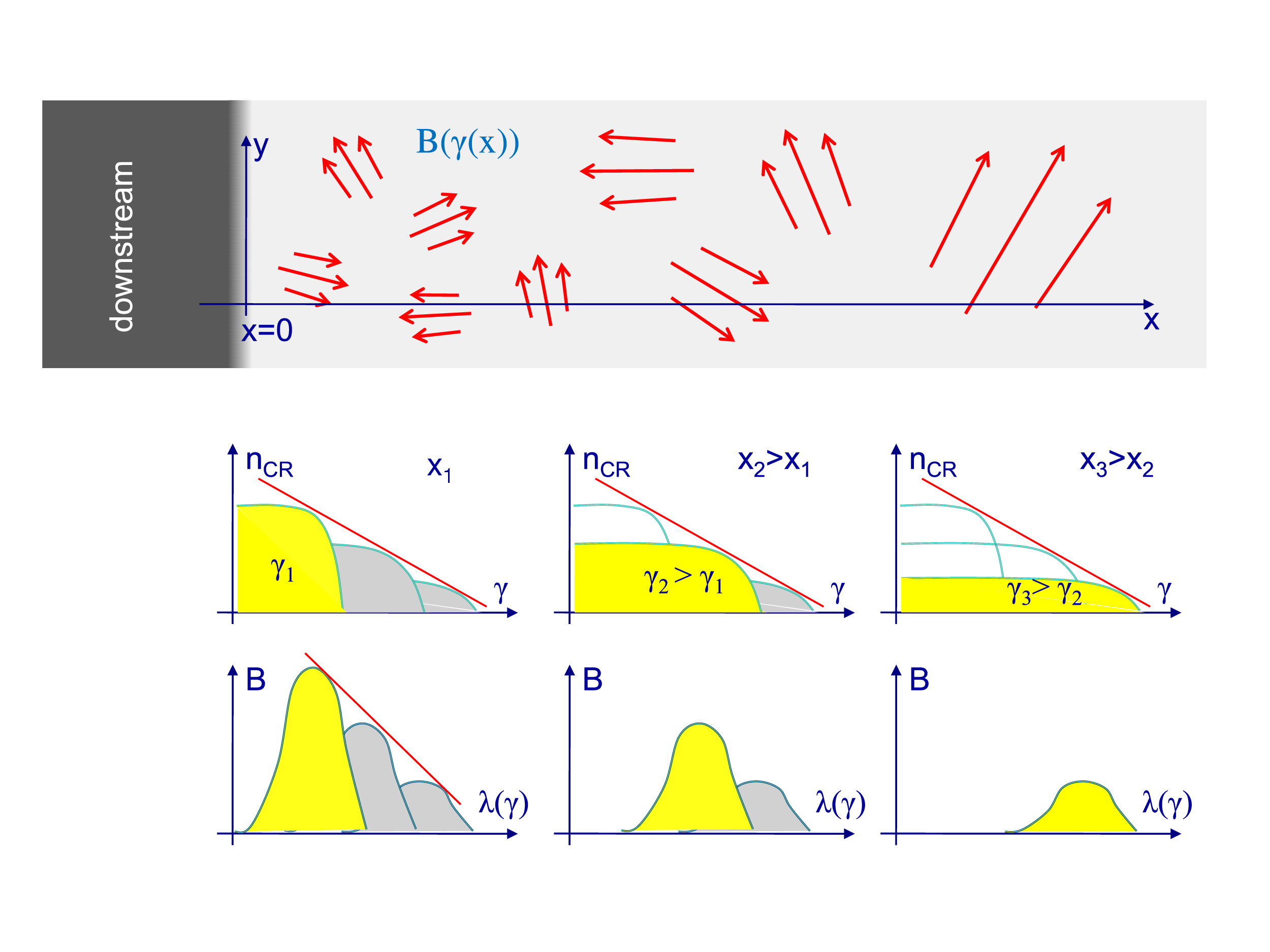}
\caption{A schematic representation of the foreshock magnetic fields: the coherence length is increasing with the upstream distance. Below are schematic graphs showing variation of the spectrum of the streaming part of cosmic rays and the corresponding self-generated fields (highlighted).
\label{f1}}
\end{figure}

These random fields deflect CR particles and ultimately lead to their isotropization. The group of particles of the Lorentz factor $\gamma_r$ thermalizes at the distance from the shock:
\beq
x_r\sim\lambda(\gamma)/(2\xi_B).
\label{x-gamma}
\eeq 
At this point, $x=x_r$, one has $\gamma=\gamma_r$ by definition; no field of the strength $B(\gamma_r)$ and the scale $\lambda(\gamma_r)$ can be produced at $x>x_r$. One can show that randomization of the higher energy particles is small in these fields at $x\sim x_r$, which means that these particles keep streaming through much larger distances $x\gg x_r$ and will produce the magnetic field further away from the shock. This field will be weaker and larger scale because of the lower density of the streaming particles $\tilde n(\gamma)\ll\tilde n(\gamma_r)$, according to Eqs. (\ref{nCR})--(\ref{lambda-gamma}). 

Finally, the number density of streaming CR particles at $\gamma_r$ is $\tilde n(\gamma_r)=n_0(\gamma_r/\gamma_0)^{-p+1}$. Therefore,
\beq
\lambda(\gamma_r)
\sim\left(m_pc^2 \gamma_0/4\pi e^2 n_0\right)^{1/2}(\gamma_r/\gamma_0)^{p/2}
\equiv\lambda_0(\gamma_r/\gamma_0)^{p/2},
\eeq 
where $\lambda_0$ is the inertial length of the lowest energy CR ``plasma". Inverting this expression yields:
\beq
\gamma_r\sim\gamma_0[\lambda(\gamma_r)/\lambda_0]^{2/p}
\sim\gamma_0(2\xi_Bx_r/\lambda_0)^{2/p}.
\label{gamma_r}
\eeq
Hereafter, the subscript ``$r$'' can be omitted without loss of clarity. 

In a steady state, this field is continuously advected toward the shock (in the shock co-moving frame since the center of mass frame of the foreshock plasma differs from the shock frame) and may affect the onset and the saturation level of the Weibel instability. In addition, the current filaments producing the fields merge with time, so that $B$ and $\lambda$ change while being advected. These nonlinear feed-back effects are difficult to properly account for in a theoretical model; hence they are omitted in the current study. PIC simulations can help us to quantify the effects as well as to confirm or disprove our assumption that the shock and the foreshock do form a self-sustained, steady state structure.

The self-similar structure of the foreshock \citep{MZ09} immediately follows from Eqs. (\ref{nCR}), (\ref{B-gamma}), (\ref{x-gamma}) and (\ref{gamma_r}). The magnetic field correlation length is proportional to the upstream distance from the shock,
\beq
\lambda(x)\sim x(2\xi_B),
\label{lambda-x}
\eeq
and its strength decreases with the distance as
\beq
B(x)\sim B_0 \left({x}/{x_0}\right)^{-(p-2)/p},
\label{B-x}
\eeq
where $B_0=\left(8\pi\xi_B m_pc^2n_0\gamma_0\right)^{1/2}$ and $x_0=\lambda_0/(2\xi_B)=\left(m_pc^2 \gamma_0/4\pi e^2 n_0\right)^{1/2}/(2\xi_B)$. In this estimate we neglected the advected fields $B(\gamma)$ as sub-dominant compared to $B(\gamma_r)$ for $\gamma>\gamma_r$. We note here that the idea of self-similarity of the Weibel turbulence has been first proposed by \citet{M+05} and then further elaborated by \citet{KKW07}, whose results are in agreement with the above scalings. The $\epsilon_B$ parameter expresses the field energy normalized to the shock kinetic energy. The energy of cosmic rays is $U_{\rm CR}=\int n(\gamma/\gamma_0)(m_pc^2\gamma)\ d(\gamma/\gamma_0)\sim m_pc^2\gamma_0n_0$ and constitutes a fraction $\xi_{\rm CR}$ of the total shock energy, $U_{\rm sh}$. The efficiency of cosmic ray acceleration, $\xi_{\rm CR}$, can be as high as several tens percent, perhaps, up to $\xi_{\rm CR}\sim0.5$, as follows from the nonlinear shock modeling \citep{V+06}. The scaling of $\epsilon_B$ is:
\beq
\epsilon_B\sim\xi_{\rm CR}\xi_B\ \left({x}/{x_0}\right)^{-2(p-2)/p}.
\label{epsilonB-x}
\eeq

These scalings hold while the shock can be treated as planar and while the ISM magnetic fields are negligible compared to the Weibel-generated fields.  If the shock is relativistic, CR particles can occupy a narrow region in front of it. Assuming CR to propagate nearly at the speed of light, their front is ahead of she shock at the distance $\Delta r'\sim\Delta r/\Gamma_{\rm sh} \sim R/(2\Gamma_{\rm sh})$ in the shock frame. Also, when the radial distance in the lab frame $\Delta r=x/\Gamma_{\rm sh}$ becomes comparable to the shock radius $\Delta r\sim R$ the curvature of the shock can no longer be neglected: the density of CR particles, which was assumed to be constant in our model, starts to fall as $\propto r^{-2}$. This leads to a steeper decline of $B$ with distance. Obviously, the first constraint is more stringent for a relativistic shock, whereas both are very similar (within a factor of two) for a non-relativistic shock. Hence one should use the first constraint. Here we also assumed that the ambient magnetic field is zero. It is a good approximation for usual ISM conditions \citep{MZ09}.

We can also estimate the magnetic field spectrum at and after the shock jump as long as dissipation is not playing a role. The magnetic field of different correlation scales created in the foreshock is advected toward the shock, so a broad spectrum is accumulated: 
\beq
B_\lambda\propto \lambda^{-(p-2)/p}\sim\lambda^{-0.091},
\label{Bspec}
\eeq
where Eqs. (\ref{lambda-x}) and (\ref{B-x}) were used and $p\sim2.2$ was assumed.

 We want to note that a number of simplifying assumptions has been made in our analysis. In particular, {\em nonlinear feedback} effects of the upstream magnetic field on the particle distribution, on the shock structure and on Fermi acceleration were omitted. The inclusion of these effects is hardly possible in any analytical model. We neglected in this model that the accelerated electrons can loose their energy and cool down in the self-generated fields while the protons keep their energy, thus causing electric fields and, possibly, currents to build up in the region and modify the foreshock structure. We also assumed that a steady state exists for the shock-foreshock system at hand. Apparently, it is not at all clear whether the steady state is at all possible or the system exhibits an intermittent behavior. One can envision a scenario in which the CRs overproduce upstream magnetic fields leading to enhanced particle scattering and the overall preheating of the ambient medium, which, in turn, can cause the shock to weaken,  disappear and then re-appear in a different place further upstream. Presently available 2D PIC simulations of an electron-position shock do show the upstream field amplification and no steady state has been achieved: both upstream and downstream fields continue to grow for the duration of the simulations though no shock re-formation has so far been observed. We argue that extensive PIC or/and hybrid simulations of a shock are imperative for further study.

\section{Conclusions}

What is more important: a shock or its foreshock?
Conventionally, one believes that the radiation observed from gamma-ray bursts is produced in the shock downstream. Indeed, the magnetic fields there are present, though the small-scale ones decay relatively fast. The foreshock fields advected into the downstream may increase the meso-scale fields which are not coupled to dissipation efficiently. The shock front itself can produce substantial radiation for certain parameters of a GRB explosion. However, the fields at the shock are mostly those created in the foreshock too and then advected with the flow. The foreshock region itself is very large in size $\sim R/\Gamma_{\rm sh}^2$ in the observer's frame and is filled with the magnetic field of sub-Gauss strength (a fraction of the CR energy density) and of correlation scale as large as $\sim R/\xi_B\sim0.01R\sim10^{16}$~cm for $R\sim10^{18}$~cm. The present model is very simple and neglects a number of important nonlinear feedback effects. Extensive simulations are needed to accurately study the foreshock structure. How much radiation is emitted from the foreshock region strongly depends on these details. It is premature to draw any conclusions on this matter. Summarizing all the above, we believe that the study (e.g., using PIC and other simulations) of a foreshock is of crucial importance for our understanding of observational signatures of relativistic shocks and it shall be given the top priority.

\begin{theacknowledgments}
The author thanks colleagues at IKI and RRC ``KI" for discussions and comments. This work has been supported by NSF grant AST-0708213, NASA ATFP grant NNX-08AL39G, and DOE grant  DE-FG02-07ER54940.
\end{theacknowledgments}

\bibliographystyle{aipproc}   

\end{document}